# Grain boundary defects induced $T_C$ increment in MnSi

## Authors

Adrian Benedit-Cardenas[a], Tobias Fox[b], Stéphanie Bruyère[a], Christoph Pauly[bc], Flavio Soldera[b], Sylvie Migot[a], Frank Mücklich[b], David Horwat[a], Alexandre Nominé[ad] a: Institut Jean Lamour, UMR CNRS 7198, Université de Lorraine, BP 70239, F-54506, Vandoeuvre-le`s-Nancy, France.
b: Chair of Functional Materials, Saarland University, Campus D3.3, Saarbrücken 66123, Germany
c: Center for Correlative Microscopy and Tomography CoMiTo, Saarland University, 66123 Saarbrücken, Germany
d: Jožef Stefan Institute, Jamova Cesta 39, SI-1000 Ljubljana, Slovenia

## Abstract

The rapid advancement of digital technologies necessitates significant progress in functional materials, which are often derived from scarce elements and involve complex manufacturing processes. Additionally, the trend towards miniaturization in high-tech devices has heightened the demand for extremely small components with tailored functionalities. In the domains of ferromagnetic materials, the market is mostly dominated by rare-earth elements-based structures, which are also limited in abundance. In this work, we focus on the microstructure and properties of MnSi. It is a ferromagnetic material with a relatively low Curie temperature ($T_C$) of 30 K. However, our study demonstrates that Tc can be increased by a factor of 4 through careful control of the crystal size. MnSi thin films were synthesized by combining two non-equilibrium techniques: magnetron sputtering and laser annealing. Laser annealing provoked the crystallinity evolution, by heat accumulation, of the barely crystallized films deposited by magnetron sputtering. The laser beam scanning parameters were adjusted to achieve different fluence values, pulse numbers, and pulse frequencies at each point of the film. Films with a crystal size of around 20 nm exhibited a $T_C$ of up to 120 K. These properties were obtained under conditions of low fluence and a high number pulse. Local laser impacts were applied to as-deposited samples, enabling spatially controlled crystallization. The interface between the poorly and well-crystallized regions was showcased using high-resolution transmission electron microscopy (HR-TEM). A spatial resolution of approximately 100 µm was achieved. These results demonstrate the strong potential of laser annealing as a versatile and promising approach for the fabrication of miniaturized devices.

## Introduction

The next generation of technology is defined by the convergence of deeper digital integration within society and the development of eco-friendly technologies. This dual transition toward greener and more digital solutions is accompanied by an important demand for raw materials, raising significant concerns about their future availability. This availability of raw materials is a determining factor in sustaining a robust high-tech materials industry. However, the growing scarcity of many essential raw materials imposes a re-evaluation of current approaches. Traditional considerations of cost and performance must now be balanced with sustainability factors, as material abundance becomes a key criterion.

Environmental impact is a crucial parameter when analysing the separation processes of elements from ores. These processes are highly energy-consuming, with energy often derived from fossil fuels, thereby contributing to increased $CO_2$ emissions in the atmosphere. Companionality is defined as the extent to which the production of an element is intrinsically tied to its role as a by-product of a host element's extraction from geological ores [1]. The production of such by-products is fundamentally subordinated to the extraction of the host elements rather than to fluctuations in market demand, making price prediction particularly challenging [2]. In this context, rock-forming elements such as copper, aluminium, iron, titanium, silicon, and manganese take centre stage due to their high abundance [3], low $CO_2$ footprint [4], and low companionality [1].

Silicon and manganese were selected for this study, with a particular focus on the MnSi compound. Its ferromagnetic behaviour arises from itinerant electrons forming a helical magnetic structure [5,6]. In this structure, spins separated by about 87 Å are antiparallel [7]. The helical magnetic ordering of MnSi is especially significant because it suggests the potential presence of skyrmions in the material [8]. However, these magnetic properties are only observable below the Curie temperature $T_C$ of approximately 30 K [9], which is well below the technological threshold of liquid nitrogen at 77 K.

Rylkov *et al.* investigated strategies to increase the Curie temperature of MnSi and identified Si sub-stoichiometry within the MnSi phase as a promising approach [10,11]. The enhancement of $T_C$ is attributed to two main factors: the formation of local magnetic moments associated with defects and the interactions between these moments. Furthermore, it was shown that the magnetic response exhibits a global rather than purely local character. Although the concentration of defects in Si sub-stoichiometric MnSi is relatively low, it leads to a significant increase in the Curie temperature. Si vacancies are responsible for this behaviour [10].

It has been proposed that the removal of Si atoms from the first-neighbour shell of Mn atoms generates dangling bonds at Mn sites. These dangling bonds induce the localization of Mn 3d-orbital electrons, resulting in the formation of magnetic defects expressed as Mn atomic moments [12]. These moments enhance the robustness of ferromagnetic coupling against thermal agitation, thereby increasing the Curie temperature. On the other hand, according to the Mn–Si phase diagram, MnSi does not naturally accommodate significant deviations from stoichiometry under stable equilibrium conditions. Even small deviation from the stable equilibrium composition can lead to thermodynamic phase separation [13].

Besides accurate choice of element, the local control of properties attracts a lot of technological interest. The ability to spatially localize a phase transition is of major technological interest because a confined structural change can induce a dramatic modification of the functional response of a material [14–16]. This makes it possible to define, within the same continuous film, regions with strongly contrasting magnetic, electronic, or optical properties, without requiring a change in global composition. Such localized control is particularly attractive for miniaturized devices, where functionality must be written with high spatial precision. Examples include data storage, where binary information relies on the creation of small regions with distinct readable states, as well as logic devices, sensors, and reconfigurable functional surfaces. In this framework, local laser-induced crystallization appears as a powerful strategy to directly pattern properties through microstructural design.

The strategy of this work is to tune the concentration of Si vacancies in the MnSi structure by controlling the grain boundary density. Since each grain boundary intrinsically introduces a concentration of dangling bonds, modifying the boundary density provides a pathway to regulate defect formation. A two-step process is proposed, consisting of film deposition by magnetron sputtering followed by controlled crystallinity evolution through laser irradiation. A similar strategy (magnetron sputtering and thermal annealing) was proposed in a previous work of our group and grain boundaries concentration was proposed as a factor that increases $T_C$ [17]. In the present work, the laser irradiation conditions determine the resulting microstructure, which in turn governs the magnetic response of the films, with grain boundaries playing a decisive role. A direct relationship is thus established between the degree of crystallinity and the magnetic behaviour of the films, leading to an increase in the Curie temperature. To this end, local crystallization is explored as a promising tool for spatially controlled tuning of the magnetic properties.

## Materials and Methods

Films were synthesised by magnetron sputtering from two independent Si and Mn targets. The deposition was made on 10 X 20 mm$^2$ glass substrate and took place inside a 40-l sputtering chamber. Prior to deposition, the chamber was pumped down using a primary mechanical and a secondary turbomolecular pumps down to a pressure of 10$^{-5}$ Pa. Argon was used as sputtering gas at a fixed mass flow of 50 sccm. The base pressure before starting deposition was around 0.35 Pa.

The substrates were placed in an azimuthal position respective to the targets (i.e 8.1 cm from the substrate holder rotation axis). To ensure coating homogeneity, the sample rotation was set to 210 rpm. The distance between the substrates and the targets was fixed to 5 cm. The targets were placed parallel to the substrate-holder at 8.1 cm distance from the substrate-holder rotation axes. The distance between targets was 14 cm. Both Si and Mn targets, Kurt J. Lesker Company manufacturing with a purity of 99.999% and 99.95%, respectively, had a diameter of 50 mm and were 3 mm thick. They were independently current-controlled. The applied currents were 0.60-0.80 A and 0.13-0.15 A for Si and Mn targets respectively. These conditions assured an equal atomic composition of silicon and manganese. The pressure during deposition was 0.39 Pa. Poorly crystallized films ware obtained.

Laser annealing was performed using a PX-series picosecond laser system, with the machine and control software supplied by Pulsar Photonics. The laser wavelength was 532 nm, with a pulse duration of 10 ps. Single-pulse laser impacts were performed on magnetron-sputtered as-deposited films. Laser irradiation can trigger several distinct mechanisms in condensed matter, including melting [18,19], boiling or swelling [20–23], burst phenomena [24], and non-thermal vaporization (ablation) [25,26]. These mechanisms significantly affect both the surface topography and the microstructure of the irradiated material. As a general trend, signatures of both thermal and non-thermal vaporization were observed, see figure S1. The modified depth tends to converge toward a maximum value of approximately 35 nm, see figure S2.

Despite these effects, single-pulse irradiation was insufficient to induce crystallization, see figure S1. Instead, crystallization requires the application of multiple laser pulses delivered sequentially. Crystallization was possible as a collective effect arising from cumulative pulse interactions rather than the simple superposition of independent single-pulse events. This effect has been already reported [27,28]. Crucially, the pulse frequency plays an important role in this phenomenon. Following each pulse, the absorbed heat initiates a process of dissipation, and the temperature experiences a decay, following an exponential pattern. The accumulation of heat will persist if the next pulse occurs before the material has been allowed to cool down to room temperature, see figure S3.

Full-film crystallization of the as-deposited films was achieved by scanning the sample surface with a pulsed laser beam. It required a careful combination of several parameters, including laser fluence, spot size, pulse repetition frequency, and scanning speed. The minimum achievable time interval between consecutive pulses was 10 µs (frequency equal to 100 kHz), while the maximum scanning speed reached 40 mm/s. In addition to these parameters, the laser fluence was further tuned by adjusting the focal conditions. Two configurations were employed, hereafter referred to as focused and non-focused. The laser spot diameter was 40 µm and 80 µm for focused and non-focused conditions, respectively. The non-focused beam distributes the laser power over a larger area, thereby reducing the fluence and enabling finer fluence control. The laser beam exhibits a circular spot profile on the surface; as a result, a rectangular area cannot be homogeneously covered. Consequently, identical irradiation conditions are only satisfied on average across the treated area. A detailed description of the scanning strategy is provided in the Supplementary Information.

A cold FEG JEOL ARM200 transmission electron microscope with 2 Cs correctors was used for conducting transmission electron microscopy (TEM) and energy dispersive X-rays spectroscopy (EDS) chemical analysis on extracted lamellae. Those lamellae were extracted from films by focused ion beam (FIB) with scanning electron microscopes (SEM) dual beam system FEI Helios Nanolab 600 and 600i. XRD was performed in a Bragg-Brentano configuration by a diffractometer Bruker D8 Advance with KαCu radiation (λ = 1.5406 Å).

A MPMS3 superconducting quantum interference device (SQUID) vibrating sample magnetometer (VSM), provided by Quantum Design Inc., was used for measure the magnetic moment of samples under chosen magnetic field H. Samples were cut in smaller parts of around 4 X 4 mm$^2$ area to this purpose. The system is equipped with compartments containing liquid nitrogen and helium. That allows measurement at temperatures T from 1.8 K to 300 K. Magnetization M was measured by establishing the magnetic field parallel to the films (in-plane, IP). The M vs T curves were measured under a constant magnetic field, $\mu_0 H$ = 0.5 T, where $\mu_0$ is the vacuum magnetic permeability. Besides the films, a reference uncoated substrate was measured alone to determine its magnetic signal. That signal was subtracted from the M vs T curve considering the substrate and sample areas.

## Laser scanning annealing

The as-deposited films exhibit crystallinity at its earliest stage. Nevertheless, laser irradiation enables a significant evolution of the film crystallinity. Under focused irradiation conditions, corresponding to fluences on the order of tenths of J/cm$^{-2}$, crystallization was achieved after the accumulation of several tens to thousands of pulses. In contrast, under non-focused conditions, where the fluence is reduced to the order of hundredths of J/cm$^{-2}$, crystallization required tens of thousands of pulses. An overview of the structural and magnetic evolution of the films as a function of the laser irradiation conditions is schematically summarized in figure 1.

Starting from the as-deposited state (figures 1a and 1d), scanning the film surface under focused conditions led to fully crystallized films characterized by a columnar microstructure (figures 1b and 1e). These columns exhibit a high degree of crystallinity. Alternatively, crystallization could also be achieved using non-focused irradiation conditions (figures 1c and 1f). In this later case, the films display a comparatively granular crystalline structure, composed of nanocrystals with characteristic sizes of several tens of nanometres, and no columnar morphology is observed. In both irradiation regimes, the formation of an oxide layer at the film surface is detected, which is attributed to annealing performed under ambient air conditions.

The magnetic properties are strongly influenced by the irradiation conditions. Films crystallized using the focused beam exhibit a lower Curie temperature, around 40 K, compared to films treated under non-focused conditions, which show a Curie temperature of approximately 120 K (figure 1g). As-deposited Si–Mn films at an early stage of crystallization, and coherence domain below 2 nm, have previously been reported to display a magnetic transition near 20 K [29,30], which has been associated with a possible spin-glass regime at temperatures below 20 K. The mechanism proposed establishes that the spin glass behaviour arises from the presence of Mn amorphous clusters.

Films become ferromagnetic with the increment of the crystal size where the Curie temperatures exceed the bulk value of 30 K. However, beyond a certain crystal size, further increment is accompanied by a decrease in the Curie temperature. A deeper understanding of this behaviour requires a more detailed investigation of the laser annealing process and its impact on the microstructure, and consequently on the magnetic properties. This is the purpose of the following investigation.

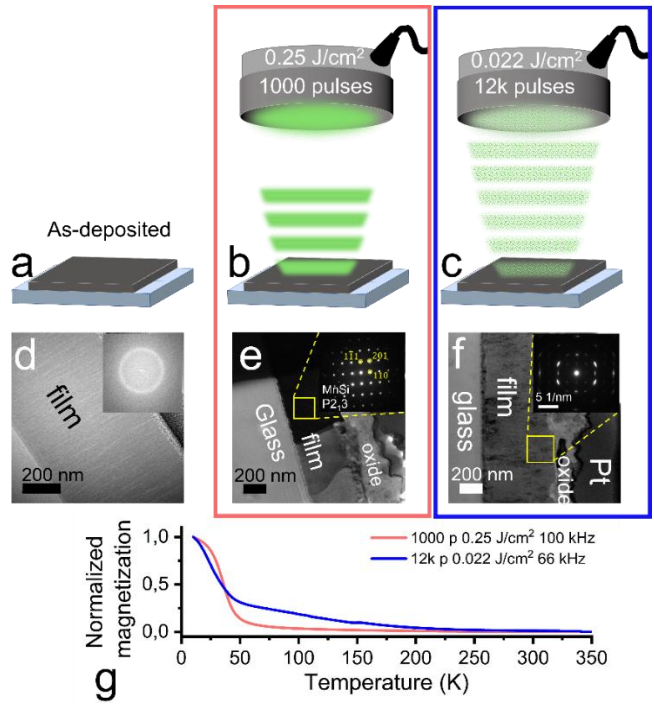

Figure 1. Schematic evolution of structure and magnetic properties with laser irradiation conditions. Films were deposited on glass substrate as represented in **a**. Two generic conditions were used to improved crystallinity by laser irradiation. The first employ a relatively high fluence (tenths of J/cm$^2$) and small number of pulses (few thousands), represented in **b**. The second employ a relatively low fluence (hundredths of J/cm$^2$) and big number of pulses (tens of thousands), represented in **c**. As-deposited films are homogeneous, as seen in TEM micrography in **d**, and at earliest stage of crystallization as shown in the FFT in the corresponding inset. High fluence irradiated films have a very well crystallized columnar structure, as seen in TEM bright field micrography in **e** and the corresponding electron diffraction pattern in the inset. Low fluence irradiated films have a crystallized granular structure, as seen in TEM bright field micrography in **f** and the corresponding electron diffraction pattern in the inset. The dependence of magnetization with the temperature at 0.5 T is shown in **g**. $T_C$ of columnar films is lower than 50 K while $T_C$ of granular films overpasses the 100 K.

Figure 2 presents the structural investigation of film crystallization based on X-ray diffractograms and TEM analysis. The diffractograms corresponding to the as-deposited film and to films annealed under different irradiation conditions are shown in figure 2a. Selected cases from figure 2a were further examined by TEM and are presented in figure 2, where each TEM image is framed with the same colour as its corresponding diffractogram to facilitate direct comparison.

The as-deposited film, represented by the black diffractogram in figure 2a and the corresponding TEM image in figure 2b, exhibits a poorly crystallized structure. This is evidenced by the presence of a very broad diffraction peak in the X-ray diffractogram and by the diffuse ring observed in the fast Fourier transform (FFT) of the TEM image, both indicative of limited short-range crystalline order.

For the focused beam configuration, the diffractogram of a film irradiated with 1000 pulses at a fluence of 0.11 J/cm$^2$ is also included in figure 2a. A comparative analysis with the film irradiated with less pulses (50 pulses) but slightly higher fluence (0.13 J/cm$^2$) reveals the existence of a crystallization threshold identified in the fluence range between 0.11 and 0.13 J/cm$^2$. Besides, it was also remarked that films irradiated at a lower frequency exhibit a lower degree of crystallization, see diffractogram of film irradiated with 1000 pulses, fluence of 0.13 J/cm$^2$ and frequency of 33 kHz in figure S5. It indicates the existence of a second critical threshold in frequency, essentials for entering the cumulative regime. These two thresholds, in fluence and in frequency, are in accordance with the accumulation heat sketched in figure S3. Nonetheless, for all films that achieved crystallization, the P2$_1$3 MnSi cubic phase is consistently observed.

At a laser fluence of 0.13 J/cm$^2$, as few as 50 pulses are sufficient to initiate the crystallization process, as evidenced by the brown diffractogram in figure 2a. A TEM analysis is presented in figure 2c and 2d. The overview image (figure 2c) reveals a homogeneous film featuring a distinct layer located near the surface, with a thickness of approximately 66 nm. The electron diffraction pattern shown in the inset of figure 2c was acquired from a relatively large region of the film where the dots correspond to a polycrystalline P2$_1$3 MnSi cubic phase.

The corresponding dark-field image, obtained using the diffraction spots highlighted in figure 2d, confirms that this near-surface layer is polycrystalline, in contrast to the underlying regions of the film. Different regions of the polycrystalline layer appear bright. Those regions are monocrystalline. The monocrystal width, in the direction of the parallel to the film surface, are generally around 100 nm. These observations lead to the conclusion that the rest of the film, apart from the polycrystalline layer, corresponds to the ring pattern in the inset of figure 2c. In other words, the majority of the film retains its as-deposited poorly crystallized state. Therefore, the crystallization induced by 50 laser pulses, at a fluence of 0.13 J/cm$^2$ and a frequency of 100 kHz, is confined to a layer on the order of tens of nanometres in thickness. This layer is responsible of the peaks corresponding to the MnSi phase in the brown diffractogram in figure 2a.

To achieve full crystallization of the film, the number of laser pulses was increased. The light-brown diffractogram in figure 2a corresponds to a film irradiated with 1000 pulses at a fluence of 0.13 J/cm$^{-2}$ and a frequency of 100 kHz. Under these conditions, the film exhibits a notably high level of crystallization. TEM analysis of this film is presented in figure 2e. The overview image reveals a columnar microstructure, with individual columns having widths in the range of 100–200 nm. The observations indicate that the columns are crystalline; however, while the crystal orientations within each column are correlated, the columns themselves are not monocrystalline.

Interestingly, the width of the columns is comparable with the monocrystals width in the polycrystalline layer in figure 2d. This coincidence points to the monocrystals, observed in figure 2d, as seeds for the growth of columns exposed in figure 2e. Crystallization expands from the surface toward the substrate as the number of pulses increases. This mechanism is believed to be responsible for the formation of the columnar structure seen in figure 2e.

Taking into account the thickness of the crystallized seed layer (66 nm), the thickness of the film (687 nm) and the number of pulses received (50 pulses) by the seed layer, and supposing a constant crystalline growth per pulse, the number of pulses necessary to crystallize the whole film can be estimated at approximately 520 pulses. Additional number of pulses would improve the crystallinity of the columns. Consequently, 1000 pulses should be more than enough to fully crystallize the film, which is confirmed in figure 2e and 2f. The electron diffraction pattern featured in the inset of figure 2e corresponds to the area selected, further confirming the well-crystallized nature of the columns, with their structure in agreement with the cubic P2$_1$3 phase of MnSi.

Interestingly, an additional foam-like layer is detectable above the film. To gain deeper insights into it an EDS mapping was conducted, see figure S6. It exhibits a notably high concentration of oxygen, providing concrete evidence of the oxide's presence. This characteristic is further affirmed by the oxygen profile line depicted in figure S6.

In figure 2a, is also presented the XRD diffractograms of films subjected to laser irradiation at a fluence of 0.25 J/cm$^2$. As expected, 50 pulses are enough for achieving crystallization, which improves with the number of pulses. In figure 2f, it is presented a TEM inspection of film irradiated with 1000 laser pulses at a fluence of 0.25 J/cm$^2$ and a frequency of 100 kHz. The initial overview image reveals a columnar film characterized by columns with increased width, in comparison with film irradiated at 0.13 J/cm$^2$, measuring between 400 nm and 500 nm. This difference in column thickness can be attributed to the doubling of the fluence. Likewise, the oxide foam that covers the film is almost twice as thick as the corresponding foam in films irradiated at a fluence of 0.13 J/cm$^2$ observed in figure 2e. The inset in figures 2f showcases electron diffraction patterns obtained from the selected area. The clear indexation of this pattern leaves no doubt about the presence of the P2$_1$3 cubic phase of MnSi within the columns.

Based on the experimental observations presented above, a crystal growth mechanism can be proposed. It has been shown that higher laser fluence leads to the formation of wider crystalline columns that growth perpendicular to the surface with the pulses. The crystallization process can be divided into two main stages. The first stage involves heat accumulation near the surface, which promotes crystal nucleation followed by lateral coalescence. The second stage is governed by heat diffusion into the film thickness, leading to oriented crystal growth propagating toward the bulk of the film.

In the first stage, the crystallization starts by crystal nucleation at different position near to the surface. This is followed by the growth in size of these coherent domains and the increase of the number of nucleation points [31]. A first fully polycrystalline near-the-surface sublayer could be composed of small crystals. A forward evolution results in bigger crystals (caused by coalescence [32,33]) as thick as the thickness of the sublayer. An example of this sublayer can be seen in figure 2d.

The thickness of the sublayer would depend on how much heat the volume close to the surface can retain before heat diffuse inside the film. Consequently, it should depend on the incoming heat per time unit (and therefore indirectly on the fluence and frequency) and the thermal conductivity. The longitudinal size of the crystals will also depend on the amount of heat accumulated before the diffusion into the film start to rule the crystallization process. After a fully crystallized sublayer, some extra heat provokes the coalescence of crystals. Therefore, wider crystals are the result of a larger amount of accumulated heat. Those crystals serve as seed for the second stage of the crystallization evolution.

In the second stage, crystallization is governed by heat diffusion and crystal growth at interfaces. Heat not retained near the surface diffuses into the film, promoting crystallization at the boundary between well- and poorly crystallized region [34]. As a result, this interface progresses from the surface toward the substrate. Initial crystal seeds grow along the temperature gradient. The final column width reflects the size of the initial seeds: for example, ~100 nm width seeds in figure 2d correspond to ~150 nm width columns in figure 2e. At higher fluence (figure 2f), columns become wider, indicating wider initial seeds.

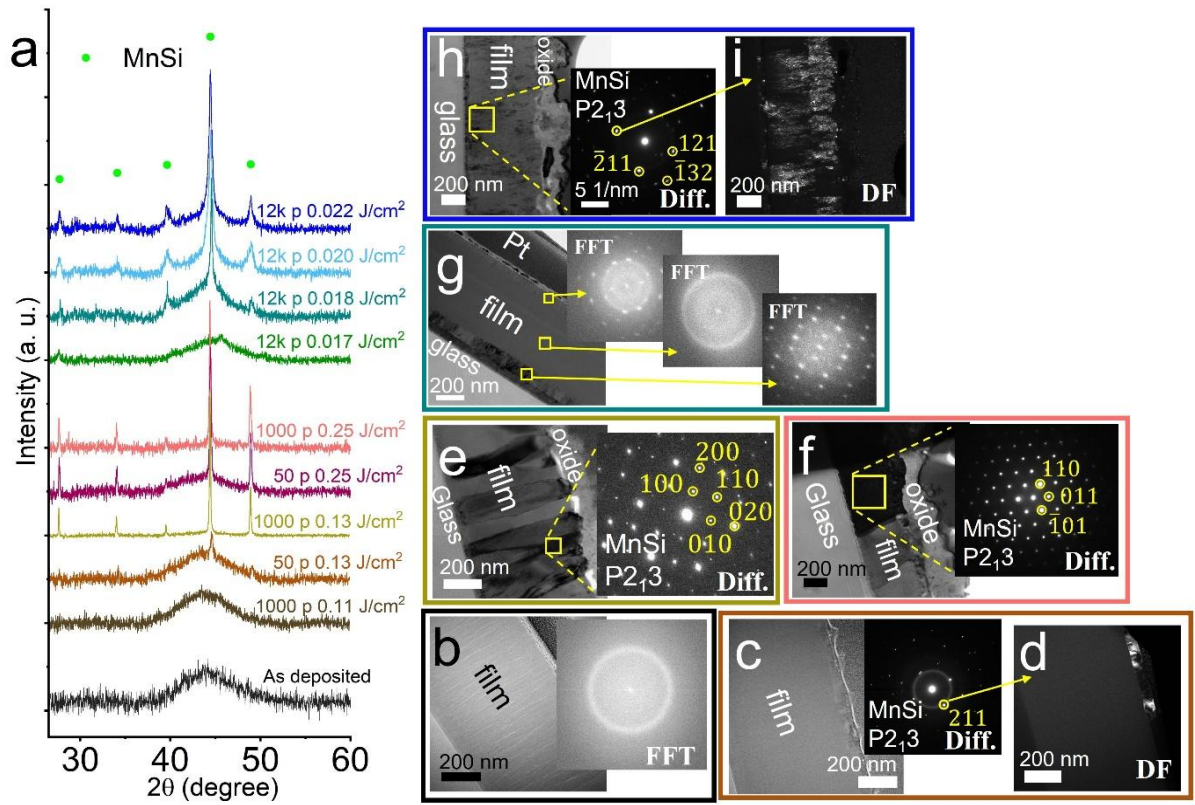

*Figure 2. Structural evolution of films under different irradiation conditions. XRD diffractograms (pulse number and fluence indicated at right) are shown in **a**. The as-deposited film shows a broad peak characteristic of an early crystallization stage, consistent with the homogeneous TEM image and diffuse FFT in **b**. Irradiation at 1000 pulses and 0.11 J/cm² yields a similar poorly crystallized structure. At 50 pulses and 0.13 J/cm², a peak at ~44° ((120) MnSi) appears, indicating partial crystallization. TEM **c** reveals a crystalline top layer above an amorphous bottom layer, confirmed by diffraction, while **d** shows oriented domains in the top layer. At 1000 pulses and 0.13 J/cm², sharp peaks corresponding to MnSi (P2$_1$3) indicate high crystallinity. TEM **e** shows a columnar structure with matching diffraction ([1$\bar{1}$1] zone axis). At 0.25 J/cm², crystallinity further improves with pulse number. Columns at 1000 pulses observed in **f** are thicker than those in **e** with the same number of pulses and lower fluence. For 12000 pulses, crystallinity increases with fluence: at 0.018 J/cm² (**g**), crystalline top and bottom layers surround a poorly crystallized middle layer unaffected by the laser, while at 0.022 J/cm² (**h**), the film becomes fully crystalline and homogeneous; **i** shows randomly oriented domains and a two-layer contrast in dark-field imaging.*

The green and blue diffractograms in figure 2a correspond to non-focused beam conditions. In these cases, the repetition frequency was fixed at 66 kHz and the number of pulses at 12000, while the fluence was varied. The number of pulses required to crystallize the films is significantly higher under non-focused conditions because the applied fluence is approximately one order of magnitude lower than for the focused beam case. The formation of the P2$_1$3 cubic phase of MnSi is confirmed, as observed for films irradiated under focused beam conditions. As anticipated, the diffractograms reveal an increase of crystallinity with the fluence. The crystallization threshold is located between fluence values of 0.017 and 0.018 J/cm².

However, films irradiated at 0.017 J/cm² can also be crystallized by increasing the repetition frequency. Figure S7 presents X-ray diffractograms of films irradiated with 12000 pulses at a fluence of 0.017 J/cm², with varying pulse frequencies. Higher repetition frequencies lead to stronger crystallization, which can be attributed to increased heat accumulation, as discussed previously. This observation highlights the critical role of pulse frequency in the crystallization process.

Figure 2g shows TEM analysis of a film irradiated at a fluence of 0.018 J/cm². The corresponding XRD diffractogram is displayed in Figure 2a, indicated by the matching frame colour. Three distinct crystalline regions are observed: near the surface, in the central part of the film, and close to the substrate. The corresponding FFT patterns for these regions are also shown in figure 2g. A granular crystalline structure dominates near both the film surface and the substrate, whereas a poorly crystallized region persists in the film centre. The structure in the central region is similar to that of the as-deposited film (figure 2b), indicating that this region remains largely unaffected by the laser under these conditions. The spatial distribution of crystalline regions suggests that crystallization propagates inward from both the surface and the substrate, in contrast to the behaviour observed under focused beam irradiation.

Figures 2h and 2i present the TEM inspection of a film irradiated at a fluence of 0.022 J/cm² with 12000 pulses at a repetition frequency of 66 kHz. The corresponding X-ray diffractogram of this film is shown in figure 2a, indicated by the matching frame colour of figures 2h and 2i. Figure 2h provides an overview of the film, with the yellow square highlighting the regions where electron diffraction was performed. The corresponding electron diffraction pattern is also shown in figure 2h. The dark-field image in figure 2i was obtained by selecting the diffraction spots indicated in the electron diffraction pattern of figure 2h.

The dark-field images reveal the presence of two distinct sublayers within the film. Both layers are crystalline and exhibit a granular microstructure, with no evidence of a poorly crystallized core at the film centre. This two-sublayer configuration represents an advanced stage of the region arrangement observed for films irradiated at a fluence of 0.018 J/cm² (see figure 2g). Films treated under non-focused laser spot conditions display a granular microstructure different to films irradiated under focused beam conditions, which typically exhibit a columnar microstructure. In addition, the presence of an oxide foam layer covering the film surface is evident in figure 2h. This oxide layer is consistent with that observed in films processed under focused beam conditions.

It can be used the same basics, as for the columnar films, to explain the crystallization process for the granular films. The critical difference is that, as fluence is one order lower, the volume close to the surface cannot accumulate heat before it propagates into the film. In such a scenario, the slow increment of input thermal energy allows a uniform distribution of temperature throughout the film. The boundaries at the surface and substrate play a pivotal role by providing the supplementary energy required for crystallization [34]. Initial crystalline sublayers near the surface and the substrate appear. These sublayers are granular polycrystalline structured, as the incoming energy per time unit is not high enough for provoking coalescence. A new moving interface is established between the granular crystallized sublayers and the poorly crystallized centre layer. That is the case of figure 2g. The film is fully crystallized when both crystalline sublayers grow and meet inside the film. That is the situation of figure 2h.

Summarising, the film crystallizes in monocrystalline columns from the surface to the substrate in focused beam regime. While for non-focused regime, film crystallize from surface and substrate to the centre in granular polycrystalline structure.

## Magnetic response

The magnetic response of films irradiated under different conditions was examined, and the results are presented in figure S8. In general, the M vs H curves exhibit negligible hysteresis, making it difficult to estimate the Curie temperature by tracking the evolution of remanent magnetization. Instead, M vs T curves were measured and are also shown in figure 3.

As discussed previously, the Curie temperature increases with the concentration of grain boundaries rising dangling bonds. This correlation arises from the formation of magnetic defects associated with a reduced Mn coordination number. The concentration of such defects is generally inhomogeneous throughout the film. Consequently, a distribution of magnetic transition temperatures can be established across the film, such that different regions undergo the ferromagnetic–paramagnetic transition at different temperatures. This behaviour is reflected in the M vs T curves in figure 3 as a quasi-linear decay that originates from quasi-continuous successive transitions of different film volumes at different temperatures. As a result, defining a unique Curie temperature is extremely challenging. In the following analysis, the temperature range over which the magnetization exhibits a linear decay is considered to correspond to the ferromagnetic regime, whereas outside this range the films are considered to be in the paramagnetic state.

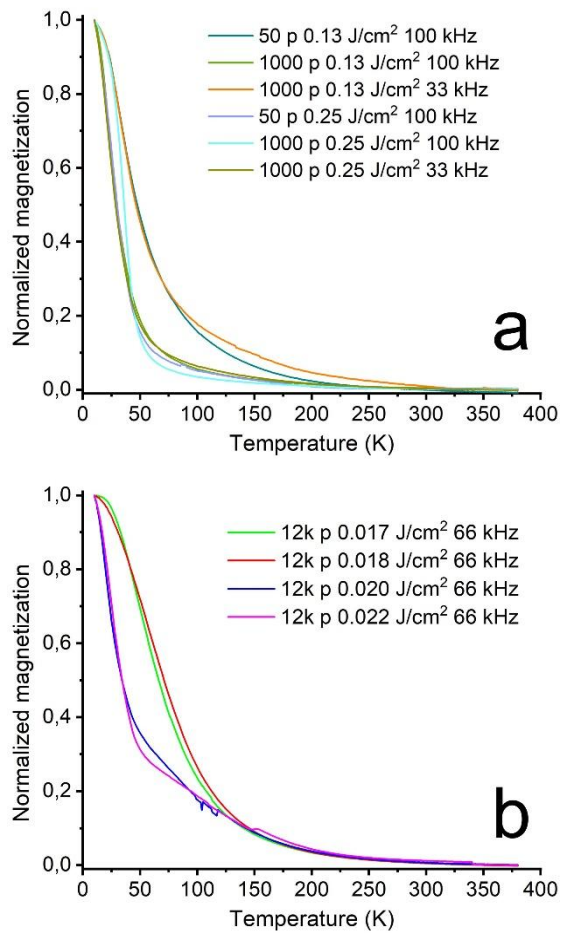

*Figure 3. Magnetic performance study of films crystallized by laser irradiation. M vs T curves of films irradiated with high and low fluences are shown in **a** and **b**, respectively. Transition occurs at lower temperatures for films irradiated with a higher fluence and frequency as a general trend.*

In figure 3a, well-crystallized films irradiated under focused beam conditions exhibit a transition temperature that does not exceed 50 K. In contrast, films irradiated under non-focused beam conditions (figure 3b) achieve transition temperatures above 50 K. Moreover, at higher fluences (0.020 and 0.022 J/cm²), the M vs T curves reveal two distinct linear decay regions: one below 50 K and another extending beyond 100 K (120 K), indicating that a transition temperature above 100 K can be reached. Films irradiated for longer durations at lower fluences demonstrate enhanced magnetic performance. Multiple observations indicate variations in crystal domain sizes, which directly influence the concentration of grain boundaries and, consequently, the magnetic behaviour.

## Relation between laser condition, structure and magnetic response of films

Figure 4a presents a comprehensive analysis of crystal size values (y-axis) obtained by applying the Scherrer equation to X-ray diffractograms, covering the full range of irradiation conditions. The x-axis represents fluence values, while the number of pulses is indicated by the colour scale (see legend in figure 4a), and different symbols correspond to distinct repetition frequencies. As a general trend, lower fluences result in smaller crystal sizes, a behaviour that persists even when a significantly higher number of pulses is applied under low-fluence conditions. This observation confirms that films irradiated at lower fluences barely exceed the threshold for the cumulative crystallization regime.

Figure 4b presents a comprehensive map illustrating the relationship between the Curie temperature and the irradiation conditions. Similar to figure 4a, the full range of irradiation parameters is represented. The highest $T_C$ values are achieved in films irradiated at low fluences with a high number of pulses.

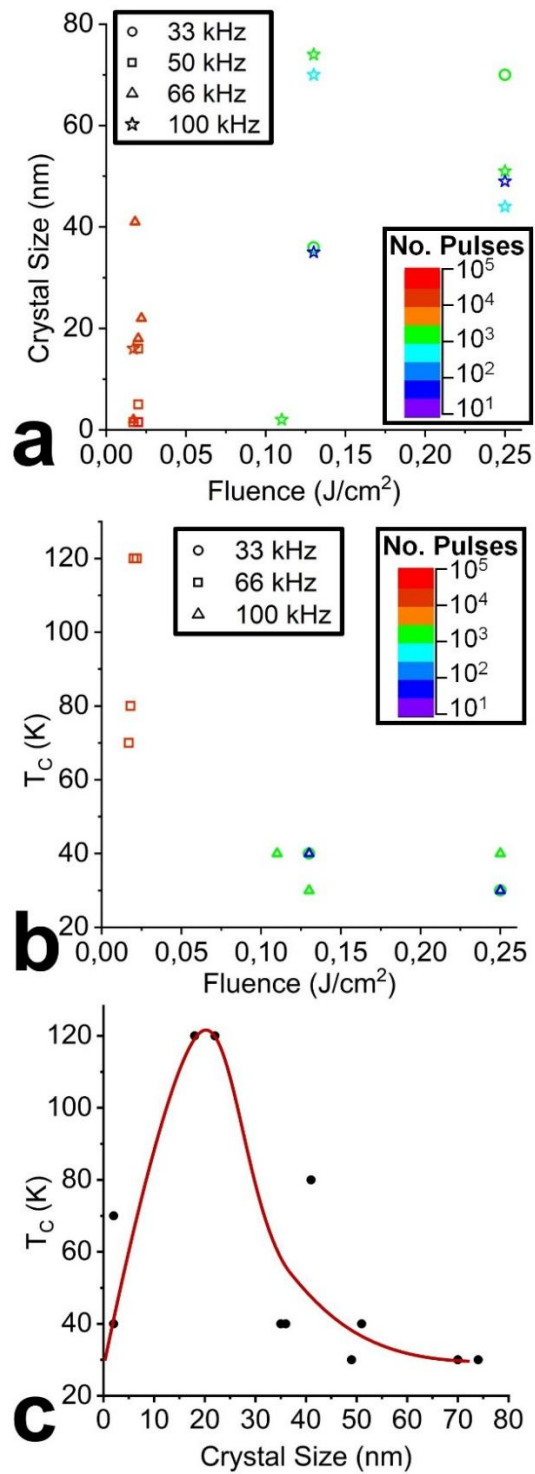

*Figure 4. Relation between the irradiation conditions, the microstructure and the magnetic response. The map in **a** depict the relation between the crystal size and the irradiation conditions. The x axis scan through all the values of fluence. The number of pulses is defined by the scale colour and the frequency values are indicated by the symbol of the scatter as shown in the legend. The map in **b** express the relation between the Curie temperature and the irradiation conditions, being the information exposed as in **a**. The lower values of crystal size correspond with the higher values of Curie temperature and are achieved by the lowers fluences as a general trend. The dependence of the Curie temperature with respect to the crystal size is shown in **c**. The optimal values of crystal size lie around 20 nm.*

Figure 4c explores the dependency of $T_C$ on crystal size. Notably, the optimal crystal size for achieving the highest $T_C$ is around 20 nm. However, this optimal point should be interpreted with caution, as some data points in figure 4c correspond to films that are only partially crystalline. In these cases, the magnetic behaviour is strongly influenced by the presence of poorly crystalline regions. As previously discussed, amorphous films exhibit a transition temperature around 20 K. Films with a high fraction of poorly crystalline structure may show a diminished or nearly undetectable linear ferromagnetic response from the crystalline volume, due to the dominant paramagnetic signal of the region at the earliest stage of crystallization. As a general trend, smaller crystal sizes correlate with higher grain density of boundaries and, therefore, higher $T_C$. However, this trend holds only up to a certain point; when domain sizes become extremely small, the overall structural integrity is disrupted. This effect is observed in ultrafine-grained, poorly crystallized films.

For films subjected to focused beam irradiation, thick columns with reduced boundary concentrations are observed, as presented in figure 2. The XRD diffractograms of these films (figure 2a) revealed a well crystallized structures with large crystals. Consequently, the transition temperature registered are low, below 50K. Films irradiated with a non-focused beam exhibited clear granular nanocrystalline behaviour, even in fully crystallized films, with no column formation. Crystal domain sizes are on the order of tens of nanometres, as calculated from the XRD diffractograms (figure 2a). Smaller crystals imply more boundaries, potentially introducing more magnetic defects. As a result, the transition temperature exceeds 50 K for partially crystallized films (irradiated at fluences of 0.017 and 0.018 J/cm², with 12000 pulses, and a frequency of 66 kHz). Fully crystallized films (irradiated at fluences of 0.020 and 0.022 J/cm²) exhibit transition temperatures exceeding 100 K.

## Local crystallization

A further exploration of the synthesis strategy (thin-film deposition by magnetron sputtering followed by crystallization induced by laser irradiation) points towards local treatment of regions of the as-deposited films. The localized nature of the laser technique makes it possible to induce spatially controlled crystallization. Figure 5 shows the TEM inspection performed at the border of a laser irradiated spot. The spot was irradiated with 12,000 pulses at a fluence of 0.025 J/cm² and a repetition frequency of 66 kHz. Figure 5a presents a SEM image of the laser impacted region. The boundary of the irradiated area is clearly visible, and a FIB lamella was prepared at this interface, as indicated by the red rectangle. The corresponding TEM image of this lamella is shown in figure 5b.

Two distinct regions can be identified: one located toward the top of the film and closer to the centre of the laser spot (right side in figure 5b), and another toward the bottom of the film and outside the spot (left side in figure 5b). The boundary between these regions is initially perpendicular to the surface and gradually curves toward the centre of the spot, becoming asymptotically parallel to the film surface. The shape of this boundary follows the radial energy distribution of the laser beam, which exhibits maximum energy at the beam centre and lower energy at the borders. That explain the weaker penetration at the borders which is connected to the lower energy there. This energy profile is also suggested by the crater morphology observed in single-pulse spots shown in figure S1, where the depth of the impacted region increases asymptotically toward the centre of the spot.

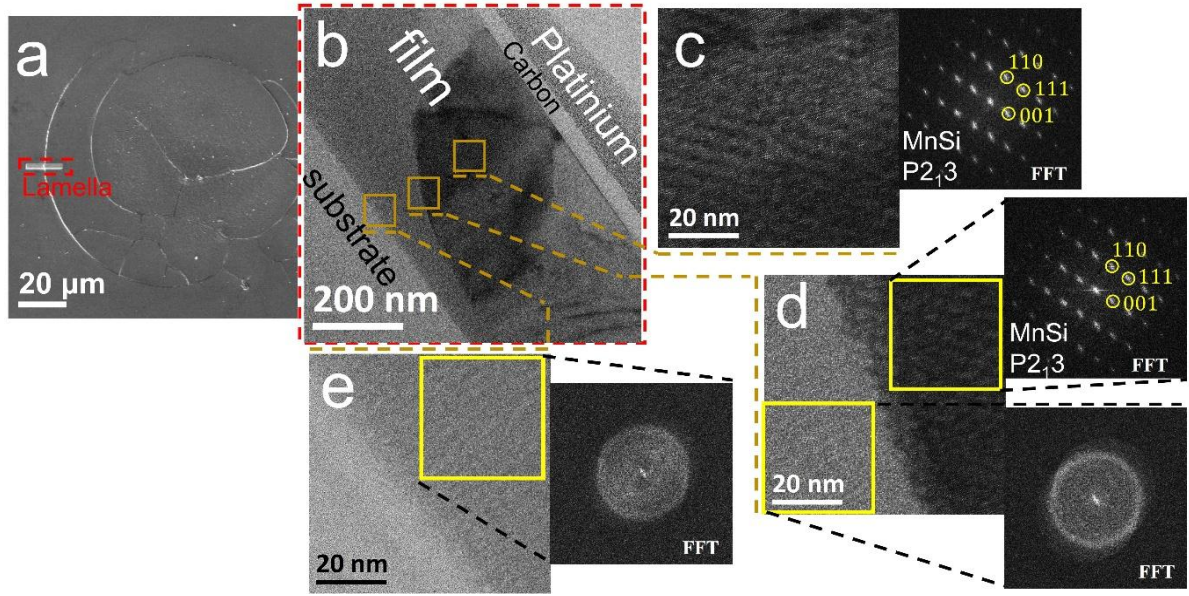

*Figure 5. Microscopic inspection of a laser impact border. The SEM micrograph in **a** shows a laser impact corresponding to 12000 pulses at a fluence of 0.025 J/cm$^2$ and frequency of 66 kHz. The red square in **a** precises the area, at the border of the impact, where the TEM lamella was taken from. TEM micrograph is shown in **b** stress out the existence of two regions in the film. Three regions were selected for high resolution transmission microscopy (HRTEM). HR micrograph in **c** shows a crystalline region, also confirmed by fast Fourier transformation (FFT). The FFT pattern was indexed as the $P2_13$ cubic structure of MnSi with ZA: $\bar{1}10$. HR micrograph in d shows the boundary between the two regions. FFT from both sides reveal a crystalline region with the same structure and orientation as in **c** on the top right of the image. At the bottom left the FFT reveals a region at the earliest state of crystallization. This region extends until the substrate as demonstrated by HR image and FFT in **e**. Laser locally crystallize the impacted area while the surroundings aren't affected.*

High-resolution TEM images and the corresponding FFT of selected regions in figure 5b are shown in figures 5c, 5d, and 5e. The laser impacted region (right side in figure 5b) exhibits good crystallinity, whereas the region closer to the substrate and outside the irradiated area (left side in figure 5b) remains poorly crystallized, similar to the as-deposited film. Fully crystallized films throughout the entire thickness are observed closer to the centre of the spot (see figure S9).

The diameter of the impacted area is approximately 100 μm; therefore, it can be established that local crystallization of the films can be achieved with a spatial resolution of about 100 μm. This resolution can be tuned by adjusting the laser focus; however, the crystallization pathway may vary, as discussed previously in this paper. The irradiation conditions (repetition frequency, number of pulses, and fluence), laser focus, final crystal structure, and achievable spatial resolution are interconnected. Consequently, targeting a specific local crystal structure requires careful optimization of all these parameters. A more detailed study of this interconnection is encouraged as a way of precisely and locally tunning properties through microstructure.

## Conclusions

In this work, it was studied the crystallization of Si-Mn films through laser irradiation. The investigation extended to understanding the magnetic performance of these crystallized films in relation to their structure. Notably, it was discovered that the crystallization process and the resulting structure were profoundly influenced by the choice of laser focus.

Scanning the film's surface, with the focused beam, led to well-crystallized films characterized by a columnar structure. TEM inspections revealed that crystallization initiated from the surface and extended toward the substrate, with column width increasing with fluence. These columns exhibited a high degree of crystallinity. On the other hand, crystallization could also be accomplished by a non-focused beam and could be enhanced by increasing fluence and number of pulses.

Crystallization, in this case, propagated from the surface and substrate towards the film's centre, revealing two distinct sublayers in fully crystallized films. These sublayers displayed a relatively disordered crystalline structure with crystal sizes of tens of nanometres, lacking the columnar structures observed in the focused beam approach.

In terms of magnetic properties, films irradiated with the focused beam exhibited the lower Curie temperature, attributed to their well-crystallized state with thick columns, resulting in fewer boundaries. In contrast, films irradiated with the non-focused spot displayed improved magnetic performance due to the high concentration of defects arising from high boundary density. It enabled films subjected to non-focused pulses to achieve higher Curie temperatures, occasionally surpassing 100 K. This approach resulted more effective in synthesizing crystalline films with an increased $T_C$. In the best scenario $T_C$ was increased by a factor of four.

Local crystallization wasn't possible by one pulse regardless the laser fluence. Only craters of different shapes, manifesting different vaporization mechanism, were obtained by single pulses. In contrast, sequential pulses were able to provoke local crystallization by heat accumulation. The spatial resolution for this local treatment was about 100 µm. This result opens a promising way of locally optimizing properties by adjusting microstructures. In that approach, the interconnection between the irradiation condition (frequency, number of pulses and fluence), the laser focus, the targeted microstructure and the spatial resolution needs to be considered.

## Acknowledgements

This work made use of the resources of the Correlative Microscopy and Tomography (CoMiTo) core facility at Saarland University.

# Grain boundary defects induced Tc increment in MnSi

# (Supplementary information)

## Authors

Adrian Benedit-Cardenas[a], Tobias Fox[b], Stéphanie Bruyère[a], Christoph Pauly[bc], Flavio Soldera[b], Sylvie Migot[a], Frank Mücklich[b], David Horwat[a], Alexandre Nominé[ad] a: Institut Jean Lamour, UMR CNRS 7198, Université de Lorraine, BP 70239, F-54506, Vandoeuvre-le`s-Nancy, France.
b: Chair of Functional Materials, Saarland University, Campus D3.3, Saarbrücken 66123, Germany
c: Center for Correlative Microscopy and Tomography CoMiTo, Saarland University, 66123 Saarbrücken, Germany
d: Jožef Stefan Institute, Jamova Cesta 39, SI-1000 Ljubljana, Slovenia

## Single-pulse laser impacts

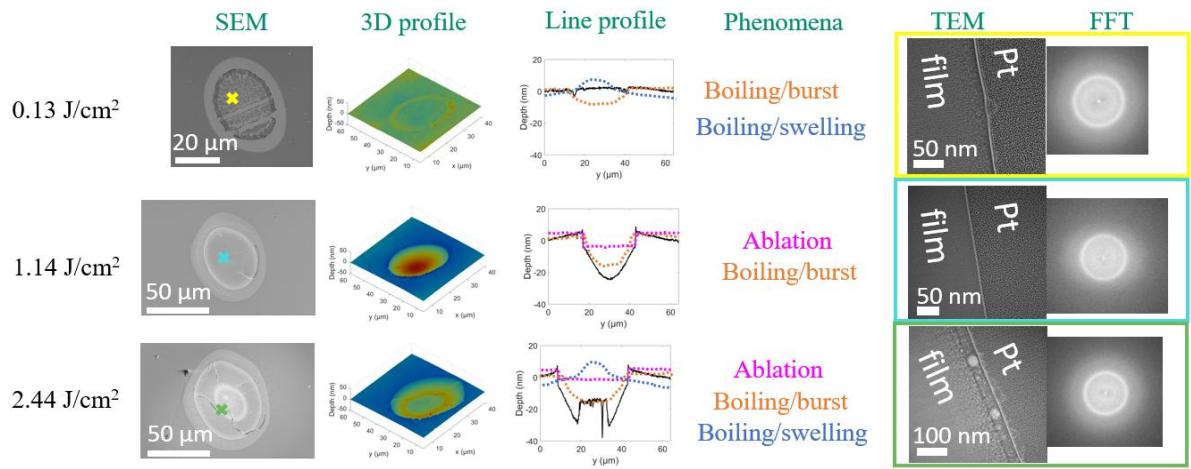

*Figure S 1. Single pulse spot inspection. Each row corresponds to the inspection to a region irradiated with the fluence expressed in the first column. SEM inspection is presented in the second column. The corresponding 3D white light interferometer (WLI) profile are shown in the third column. For a better visualization, a profile line, passing through the centre of the impact, parallel to the y axis, are shown in the fourth column. The phenomena observed in the corresponding profile are presented in the fifth column. The colour of the labelled phenomena and the profile in the fourth column match. Sixth and seventh column are dedicated to the TEM inspection and FFT.*

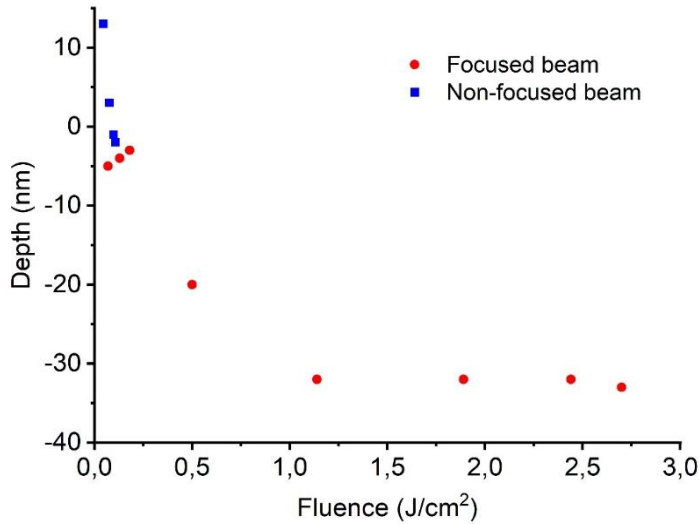

*Figure S 2. The evolution of the depth of craters, produced by single-pulse laser impacts, with the fluence, for focused and non-focused beam. The positive values of depth correspond to situation where the swelling overcompensate the burst and ablation, and a mound is formed.*

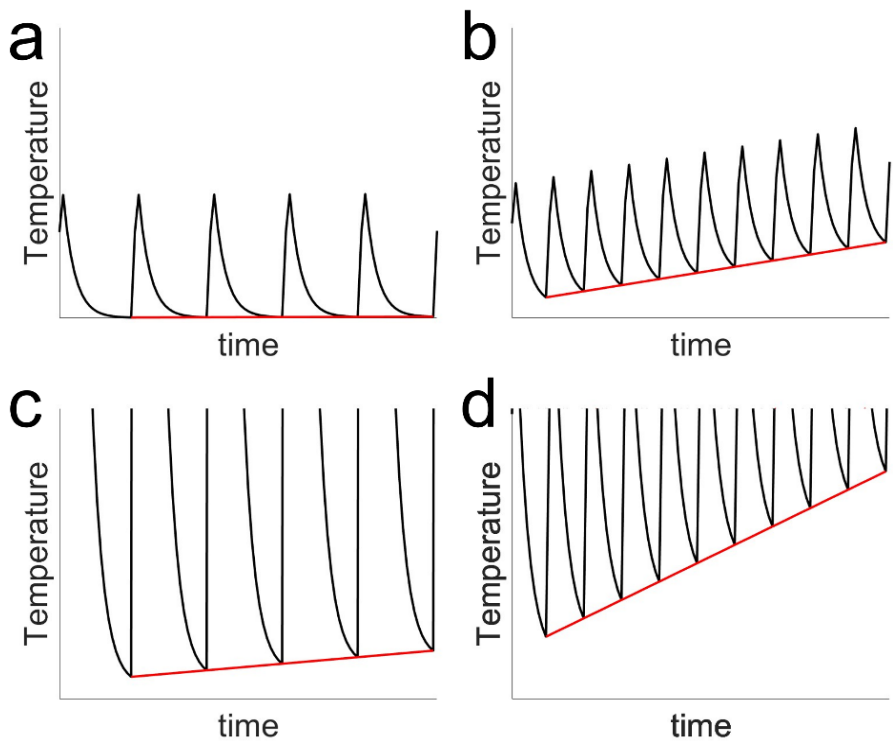

*Figure S 3. Accumulative laser heating, there are shown the four basic situations when heating a material with a pulsed laser irradiation. Situation are switched by varying the fluence and the frequency. In all the plots, the black line represents the evolution of the temperature taking into account the pulse. Temperature experiences a peak when pulses arrive. The red line represents the increment of temperature as result of the heat accumulation. Low frequency and fluence in **a** can lead to no heat accumulation. The increment of the frequency at the same fluence in **b** can lead to the cumulative regime. Likewise, the increment of the fluence at the same frequency in **c** can lead to the cumulative regime as well. The cumulative regime is assured for high frequencies and fluence in **d**.*

## Laser scanning annealing

One static spot can only crystallize a small area. The entire sample crystallization can only be achieved by moving this spot over the whole sample. The scanning speed is the speed the spot scans the whole sample surface with. Frequency pulse and scan speed should be chosen carefully to ensure equal treatment conditions in all the points of the film surface. The laser spot usually has a circular shape. A rectangular area cannot be filled homogeneously with circles, some overlapping is necessary. Therefore, the equal treatment condition is only on average. During scanning, the spot scans in the horizontal ($X$) direction, making a scanning line at first. Then, the vertical ($Y$) position is changed on step, $\Delta Y$, and a new scanning line is performed. This process is repeated until the whole area is covered, see figure S4. The quantity of pulses $n_p$ received by a point on the surface is equal to the times this point was on the surface impacted by the laser during the scanning. During the horizontal scanning, the quantity of pulses impacting each point can be estimated by:

$$n_p = \frac{D_{spot}}{\Delta X} \rightarrow \Delta X = \frac{D_{spot}}{n_p} \qquad (S1)$$

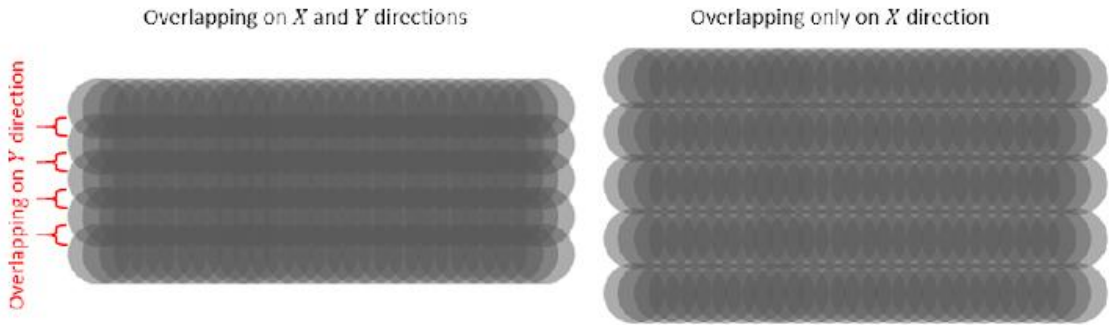

Figure S 4. Overlapping during scanning

Where $D_{spot}$ is the spot size and $\Delta X$ is the linear step between pulses. The time between pulses will not be constant and the number of pulses will be different if additional pulses from scanning lines below the original one, see figure S4, are received. In that case the analysis of the microstructure in relation with the laser parameters will be more complicated. Each point should receive all the pulses during the scanning in the $X$ direction. It is avoided to use the overlapping in the $Y$ direction, see figure S4. The step in the $Y$ direction is chosen to be 90% of the spot diameter. This value is chosen to avoid overlapping in the $Y$ direction, as well as, prevent leaving small regions between scanning lines.

$$\Delta Y = \frac{9}{10} D_{spot} \qquad (S2)$$

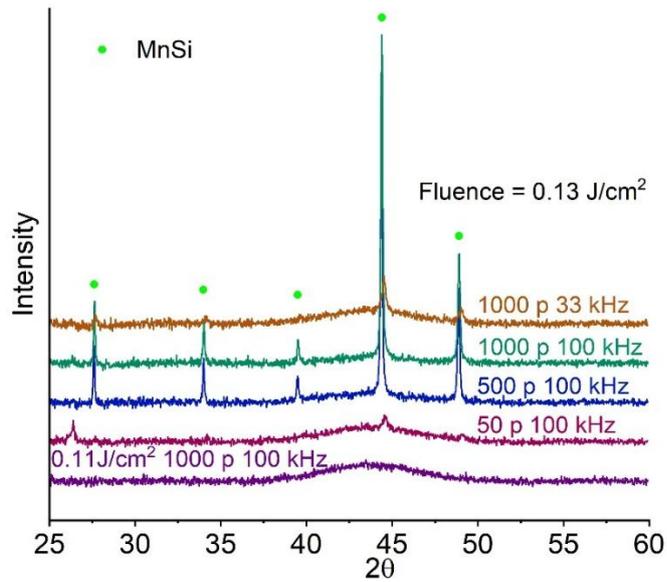

*Figure S 5. XRD of films crystallized by laser irradiation using a focused spot with a fluence of 0.13 J/cm². For comparison, it is included the diffractogram of a film irradiated with 1000 pulses at a slightly lower fluence of 0.11 J/cm². The values of frequency and number of pulses are indicated in each case. In all diffractogram, MnSi is the only phase obtained. The threshold fluence for starting crystallization lies between 0.11 J/cm² and 0.13 J/cm². Crystallization improves with the number of pulses but can be quenched if the frequency of pulses decreases.*

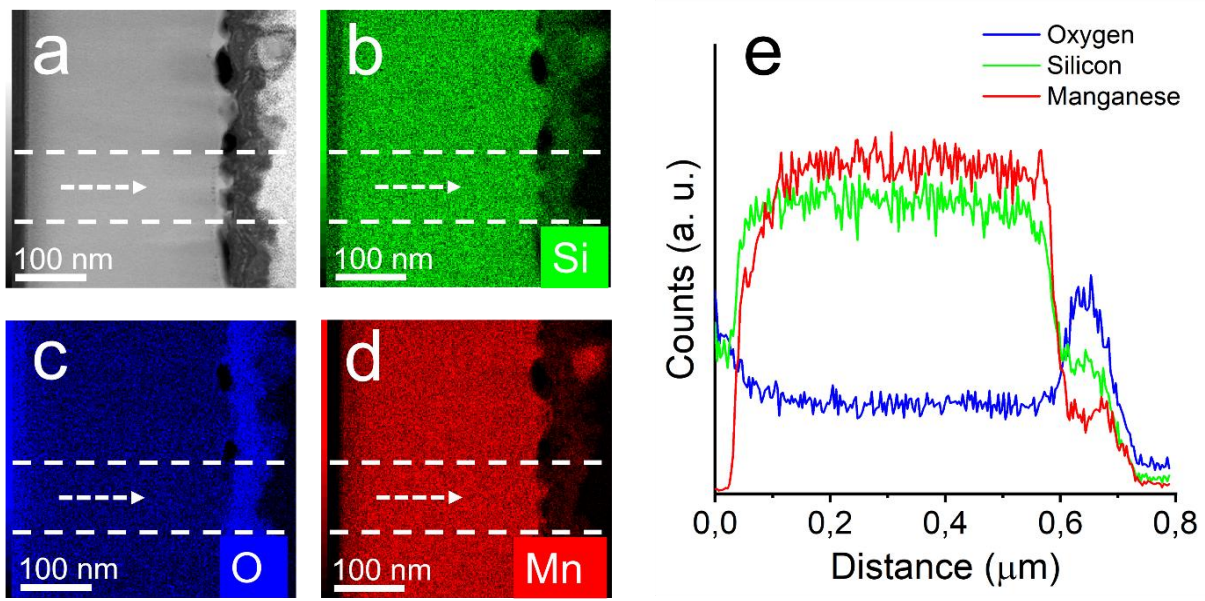

*Figure S 6. EDS cartography of an overview of film irradiated with a fluence of 0.13 J/cm², 1000 pulses and a frequency of 100 KHz. Bright field image is shown in **a**. The corresponding cartography for silicon, oxygen and manganese are shown in **b**, **c** and **d** respectively. The profile of composition in **e** corresponds to the area delimited by white discontinuous lines in **a, b, c** and **d**. The composition is homogenous along the thickness of the film. A 100 nm thick oxide layer growth on the surface of the film. The holes between the films and the oxide layer suggest an evolution toward the detaching of the latter.*

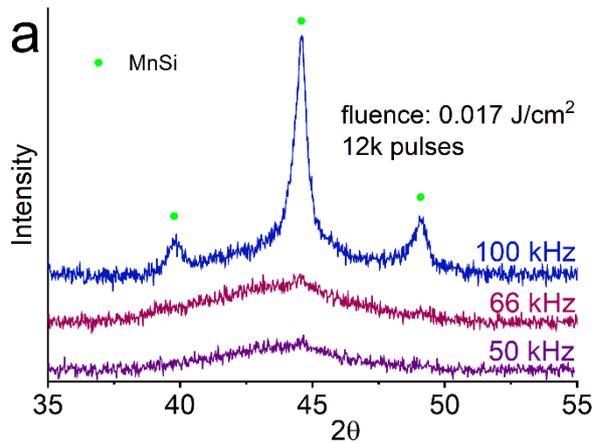

*Figure S 7. XRD diffractograms of films irradiated with a non-focused spot. Films irradiated with 12000 pulses of laser irradiation at a fluence of 0.017 J/cm$^2$. The frequency of pulses is indicated for each diffractogram.*

## Magnetic response

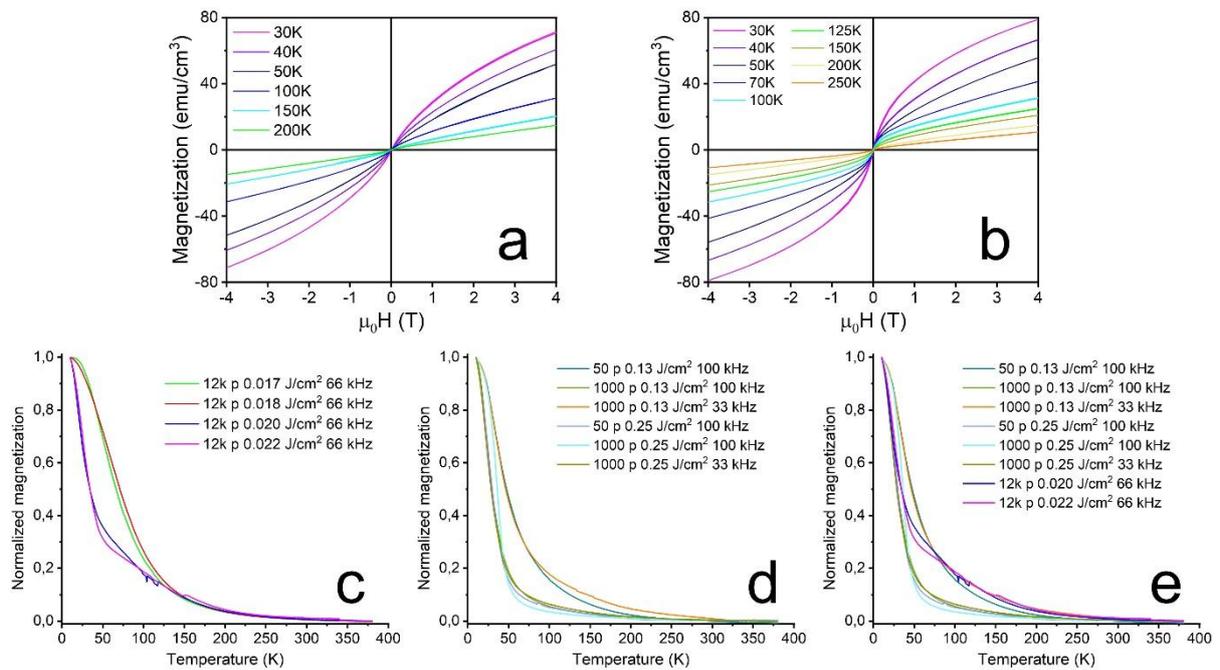

*Figure S 8. Magnetic performance study of films crystallized by laser irradiation. M vs H cycles of films irradiated with fluences of 0.020 and 0.022 J/cm$^2$, 12000 pulses and a frequency of 66 kHz are shown in **a** and **b** respectively. The cycles do not show hysteresis. M vs T curves of films irradiated with low and high fluences are shown in **c** and **d** respectively. For sake of comparison, curves in **e** correspond to both range of fluences. Transition occurs at lower temperatures for films irradiated with a higher fluence and frequency as a general trend.*

## Local crystallization

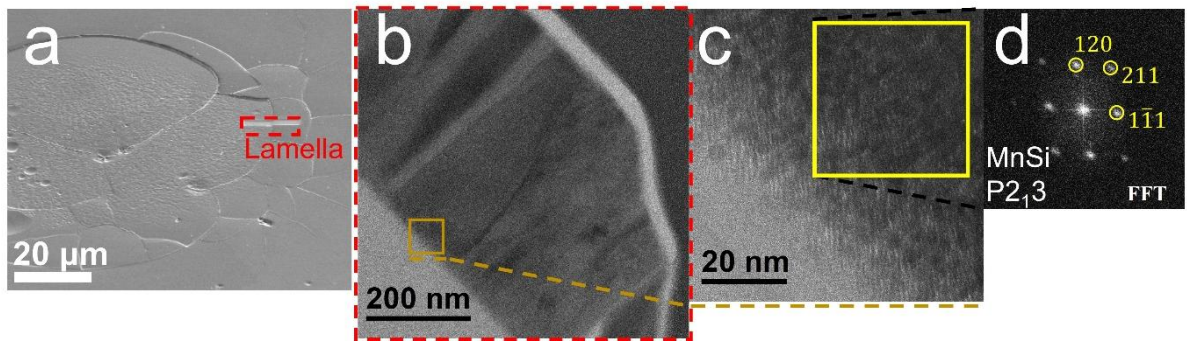

Figure S 9. Microscopic inspection on a laser impact. The SEM micrograph in **a** shows a laser impact corresponding to 12000 pulses at a fluence of 0.025 J/cm$^2$ and frequency of 66 kHz. The red square in **a** precises the area, where the TEM lamella was taken from. TEM micrograph is shown in **b** shows a general view. The framed region was selected for high resolution transmission microscopy (HRTEM). HR micrograph in **c** shows a crystalline region, confirmed by fast Fourier transformation (FFT) in **d**. The FFT pattern was indexed as the P2$_1$3 cubic structure of MnSi with ZA: $1\bar{2}\bar{3}$.